# Large magnetoresistance observed in α-Sn/InSb heterostructures

*Yuanfeng Ding,*[†] *Huanhuan Song,*[†] *Junwei Huang,*[†] *Jinshan Yao,*[†] *Yu Gu,*[†] *Lian Wei,*[†] *Yu Deng,*[†] *Hongtao Yuan,*[†,§,∥] *, Hong Lu\*,*[†,§,∥] *and Yan-Feng Chen*[†]

[†]National Laboratory of Solid State Microstructures & Department of Materials Science and Engineering, College of Engineering and Applied Sciences, Nanjing University, Nanjing 210093, China

[§]Jiangsu Key Laboratory of Artificial Functional Materials, Nanjing University, Nanjing 210093, China

[∥]Collaborative Innovation Center of Advanced Microstructures, Nanjing University, Nanjing 210093, China

\* Authors to whom correspondence should be addressed: hlu@nju.edu.cn

**Abstract**

In this study, we report the epitaxial growth of a series of α-Sn films on InSb substrate by molecular beam epitaxy (MBE) with thickness varying from 10 nm to 400 nm. High qualities of the α-Sn films are confirmed. An enhanced large magnetoresistance (MR) over 450,000% has been observed compared to that of the bare InSb substrate. Thickness, angle and temperature dependent MR are used to demonstrate the effects of α-Sn films on the electrical transport properties.

Transport phenomena in materials have been extensively investigated in history and outstanding electrical properties are always desired for electronic applications. Novel transport properties in topological materials have been emerging in recent years, such as large quantum magnetoresistance (MR),[1,2] chiral anomaly induced negative MR,[3,4] topological superconductivity,[5,6] various quantum Hall effects,[7,8] and high-performance thermoelectric properties.[9] Alpha tin (α-Sn), also known as gray tin, is an allotrope of tin with a diamond structure. It has been predicted to possess various topological phases and confirmed by angle-resolved photoemission spectroscopy (ARPES). However, its electrical properties are seldom reported till now. α-Sn (gray tin), an elemental material, is more easier for preparation and study due to its simple diamond structure, and it is predicted to hold topologically nontrivial phases under uniaxial strain.[10-12] Especially stanene, two-dimensional form of α-Sn, is predicted to be a large-gap two dimensional insulator.[13-18] However, the spontaneous transformation into β-Sn at 13.2 °C restricted the investigations on α-Sn to low temperatures before 1980s.[19-23] Farrow et al.[24] realized the epitaxial α-Sn films by molecular beam epitaxy (MBE) and the phase transition temperature was raised to 70 °C due to the substrate-stabilization effect. While α-Sn was proposed to apply for optoelectronics in view of its direct-bandgap,[25-27] it ran into a bottleneck because it is metastable. Theoretical predications[10-12] and experimental investigations[28-31] on topologically nontrivial band structures of α-Sn are reviving people's interest again. Large MR, negative MR and Shubnikov de-Haas (SdH) oscillations with nontrivial Berry phase are expected in α-Sn.[11] Compared to the band structure, however, experimental results of these transport properties have been seldom reported.

InSb is the most used substrate for α-Sn epitaxial growth due to its small lattice mismatch with α-Sn, and strain is naturally introduced by this mismatch. On the other hand, the conductivity of the InSb substrate is the main factor that brings troubles to the electrical characterization of the α-Sn films. InSb is a narrow-gap semiconductor with a high mobility, so it is also considered to have some extraordinary transport properties. Electrical properties drastically depend on the details of the InSb sample. Indeed, a large linear MR has been found in InSb under a doped or disordered condition,[32] while the MR is much lower in a high-purity single-crystalline sample.[33] And there may even exist a magnetic-field-induced metal-to-insulator transition at low temperatures and therefore negligible electrical contributions.[34] To obtain the pure properties of α-Sn, a specially designed structure has been proposed to reduce the InSb contribution,[35] and CdTe, a more insulating material, was also used as the substrate.[36] Though a more careful growth is needed, the progress has been made.

In this work, we have grown a series of α-Sn films with different thicknesses from 10 nm to 400 nm on unintentionally doped (UID) InSb substrates by MBE. The high quality of α-Sn films has been confirmed by various structural characterizations. We have taken a different perspective of showing the unique transport properties of α-Sn/InSb heterostructure as a whole and analyzing the α-Sn contributions. Despite of the significant contribution from the InSb substrate at zero field, we have observed an enhanced large MR over 450,000% compared to that of a bare InSb substrate and some attempts have been made to show its anomaly. The mix of multiple electrical conducting channels is believed to lead to the complex MR behaviors.

The α-Sn samples were grown in a group-V MBE system (Dr. Eberl MBE-Komponenten Octoplus 300) equipped with high-purity Sn (6N) solid-state sources. Figure 1a shows the sample layer structure. The unintentionally doped InSb(001) (a=6.480 Å) is chosen as the substrate for its similar crystal structure and lattice constant to that of α-Sn (a=6.489 Å). Another III-V MBE system (Veeco GENxplor) equipped with In (7N) and Sb (7N5) solid-state sources was used in this study to achieve better substrate treatment. After annealing at 350 °C in the preparation chamber, the InSb substrate was transferred to the growth chamber and heated to near 500 °C (thermocouple temperature) under the Sb protection atmosphere. The typical (2×4) surface reconstruction observed from reflection high energy electron diffraction (RHEED) patterns indicated the oxide-free InSb surface as shown in Figure 1b. After the oxide desorption, a 50 nm InSb buffer layer was grown at 450 °C to improve the surface quality. Before this template was transferred to the group-V system for α-Sn growth, an amorphous Sb layer was deposited on the InSb surface at low temperatures to avoid surface oxidation, which was desorbed near 400 °C afterwards before α-Sn growth. All the α-Sn films in this study were grown below room temperature with a background pressure < $5 \times 10^{-10}$ mbar to provide the preferable conditions for α phase formation. Correspondingly, relatively low growth rates < 0.28 Å/s were used to ensure the layer-by-layer growth. During the growth, the surface reconstruction changed from (2×4) of InSb to (2×2) of α-Sn as shown in Figure 1c.

The typical surface morphology is shown in atomic force microscope (AFM) and scanning electron microscope (SEM) images in Figure 2a and 2b. The surface roughness is rather small and below 5 nm for the 100 nm α-Sn sample. The ridge-like structure along <110> is common for the α-Sn samples in this study, which we attribute to the similar surface morphology of a clean InSb surface and this reflects the good coherence between the InSb substrates and the α-Sn films . This feature is different from that of β-Sn whose surface shows smooth and continuous structures as shown in Figure S1. The holes may be caused by the different thermal expansion coefficients between α-Sn and InSb. The high-quality α phase of the films was confirmed by X-ray diffraction (XRD)[37] and Raman spectra in Figure 2c. The α-Sn signal can be clearly observed. Figure 2d shows the cross-sectional high-angle annular dark field-scanning transmission electron microscopy (HAADF-STEM) image and corresponding energy dispersive spectroscopy (STEM-EDS) image of a 400 nm α-Sn/InSb heterostructure. We can see that the real film thickness is consistent with the nominal thickness of 400 nm. The uniform distribution of Sn signal with smooth surface and interface indicates the film is homogeneous. Weak signals from the buffer layer is due to the limited resolution considering the very close characteristic wavelengths of In and Sb to that of Sn. The thermal stability of the α-Sn films were examined by temperature dependent XRD and Raman. All the films are structurally stable under 70 °C,[37] which is important for the subsequent transport experiments. The enhanced thermal stability is mainly attributed to the substrate-stabilization effect through the coherent interface as indicated by the reciprocal space mapping (RSM) in Figure 2e. The same in-plane lattice constants of the α-Sn and InSb indicate the α-Sn films are fully strained. More structural characterizations and details about the enhanced thermal stability can be found in ref. 37.

The electrical measurements were conducted with a four-probe configuration as shown in the inset of Figure 3a in a physical property measurement system (PPMS, Quantum Design, $T_{min}$=2 K and $B_{max}$=9 T) and an integrated superconducting magnet system with variable temperature inserts (TeslatronPT, Oxford instrument, $T_{min}$=1.5 K and $B_{max}$=14 T). The indium or silver paste was used as the contacts connecting the sample surface and the holder. All the samples in our study show similar electrical transport behaviors, so the MR properties will be demonstrated by a 20 nm α-Sn/InSb heterostructure unless otherwise indicated. Figure 3a shows the temperature dependent electrical resistance of the 20 nm α-Sn/InSb heterostructure with different magnetic fields. The α-Sn/InSb heterostructure and the InSb substrate shows similar behaviors as temperature decreases at zero field, indicating a dominant contribution from the InSb substrate. However, the resistance of the α-Sn/InSb heterostructure increases drastically with increasing magnetic fields over the whole temperature range. Figure 3b illustrates the magnetic-field dependent resistance of the same sample at 2 K with the InSb substrate as a reference. MR is expressed by the relative change in resistance as $\Delta R_{xx}(B)/R_{xx}(0)\times 100\%$. A much larger MR than that of the InSb substrate was observed, reaching over 20,000% at 9 T. There are several oscillation peaks on the MR curves as indicated by the arrows in Figure 3b. The α-Sn/InSb heterostructure has clearer peaks appearing at different magnetic fields compared to the InSb substrate. However, at present we have not found any periodicity in these oscillations, neither the $1/B$ periodicity for SdH oscillations nor the $\log B$ periodicity for the discrete scale invariance in Dirac semimetals[38,39] as shown in Figure S2. This indicates that the oscillations may not be simply explained by quantized Landau levels. These features demonstrate the effects of the α-Sn films on the MR properties.

We compare the MR of the α-Sn/InSb heterostructure with different film thicknesses in Figure 3c. No obvious change in the MR magnitude can be observed, however, the intensity of the oscillation peaks reduce for the thicker films. This means that the large MR and oscillations should have different underlying mechanisms, which is further proved in Figure 3d. To demonstrate the dimensional character, angle dependent MR is measured by rotating the sample in the magnetic field as shown in Figure 3d. The large MR is suppressed with increasing angle $\theta$ between the magnetic field and the sample surface normal. The MR measured at 9 T scales linearly with $\cos\theta$ as shown in the inset of Figure 3d. This anisotropic character corresponds to a quasi-2D Fermi surface. On the other hand, the oscillation peaks do not shift with $\theta$, indicating a 3D feature. Since the oscillations of a 2D system would only depends on the normal component of the magnetic field, they will appear at the same value of $B\cos\theta$,[40,41] which is not the case in Figure 3d. Considering the film thickness and the topological properties of α-Sn, we pointed out that the α-Sn films may introduce 2D components in the α-Sn/InSb heterostructure which enhance the MR significantly. Altering film thickness can change the dimensionality of the band structure,[34,42] induce quantum size effect,[33,43] or affect the electrical field penetrating into the substrate which may influence the oscillation intensity. According to the research of Song et al.,[42] there will be a 3D-2D transition with decreasing thickness within several tens of nanometers. This may be not the case in this study as the 10 nm and 30 nm samples show similar angle dependent MR as in Figure S3. Therefore the oscillations may essentially come from the InSb bulk. However, the surface and interface states will survive as long as the film is thick enough not to hybridize them, and typically, 10 nm is enough.[34] This is in consistence with the relatively constant MR magnitude with different film thicknesses and its 2D feature. Moreover, Two-dimensional electron gas (2DEG) can form at the α-Sn/InSb interface, and may contribute to the electrical transport.[44,45] Therefore, the complex MR properties may be caused by the multiple conducting channels in the α-Sn/InSb heterostructure. Most surprisingly, it is unusual to see an obvious change in MR behavior since the InSb substrate should dominates the electrical transport. Because of the enhanced MR compared to the InSb substrate, it further indicates that the MR properties are not only from α-Sn. Generally, the shorting effect of the InSb substrate will shadow the intrinsic electrical properties of the α-Sn films, however, it may be reasonable to consider the effects of the α-Sn on the properties of InSb. We assume some possible mechanisms that may be related to the enhanced MR. Since both the α-Sn and InSb contribute to the electrical transport, the charge carrier transfer between the two is also involved. This remind us of the possible potential barrier at the interface. The barrier can reduce the transport in the InSb bulk but in turn increase the contribution from the α-Sn films and the interface. If these conduction channels and/or the barrier show a different magnetic field dependence, it may contribute to the larger MR. One example is the charge resonance model widely applied for the large MR in semimetals concerning compensation between holes and electrons,[46,47] as the α-Sn films usually have intrinsic p-type carriers [28,31,35,48] while the InSb is n-type. As demonstrated below by the temperature dependent measurements in Figure 4, the thermal excitation of the carriers in the InSb substrate near room temperature can significantly suppress the large MR, which can also be attributed to the breaking of the compensation condition. Another possible way to affect the InSb properties is the modulation of the Fermi surface by the α-Sn films. In order to verify the above assumptions, the vertical transport measurement and the doping experiments are necessary, and different doping levels in those InSb substrates and α-Sn in literatures may explain why the large MR has seldom been reported before. Whether there are other transport mechanisms relevant in MR properties need further examinations.

    We have measured the temperature dependent MR as shown in Figure 4. The MR of the α-Sn/InSb heterostructure increases drastically above 9 T at 1.5 K, and reaches an extremely large value over 450,000% at 14 T as shown in the inset of Figure 4a. The

quantum extreme model proposed by Abrikosov[49,50] may be applied under these conditions. The longitudinal resistivity can be expressed as

$$\rho_{xx} = \frac{1}{2\pi}\left(\frac{e^2}{\epsilon_\infty v}\right)^2 ln\epsilon_\infty \frac{N_i}{ecn_0^2} H \qquad (1)$$

Where $\epsilon_\infty$ is the background dielectric constant, $v$ is the band velocity, $N_i$ is the density of scattering centers, and $n_0$ is the net electron density. All the electrons fall into the lowest Landau level (LL) when the temperature is low enough and the field is high enough. The oscillation intensities decrease with increasing temperature and the peaks diminish above 30 K due to the thermal blurring. However, the peak around 2.5 T shifts to higher magnetic fields as indicated in Figure 4a, which cannot be explained simply by LL splitting since temperature only broadens the levels without shifting their positions. The MR increases to a maximum around 150 K and afterwards starts to decrease. This is different from the typical trend where the MR decreases monotonically as the temperature increases.[2,51,52] Generally, the MR involves both quadratic and linear components. As shown in Figure 4a and 4b, instead of a smooth transition, there is a kink between the quadratic MR at lower field regime and the linear component at higher field regime. We introduce a negative MR component also scaling with $B^2$ to fit this transition regime as indicated by the derivative of MR in Figure S4. We note that this negative component should not be related to chiral anomaly since a dependence on the relative orientation between the current and magnetic field has not been observed, and, for example, it may be related to the weak localization effect. Therefore the MR can be expressed as

$$MR = \begin{cases} a_1 B^2, & \text{(lower field regime)} \\ a_1 B^2 + a_2 B^2 + b_1 B + c_1, & \text{(transition regime)} \\ b_2 B + c_2. & \text{(higher field regime)} \end{cases} \quad (2)$$

Where $a_1$, $a_2$ and $b_2$ are the coefficients of the quadratic MR, negative MR and linear MR, respectively, with $a_2 < 0$ and $a_1 \propto \mu_H^2$ with the Hall mobility $\mu_H$, $b_1 B$ is a linear component introduced in the transition regime, and $c_1$ and $c_2$ are constants. A fitting example is shown in Figure 4b. The lower field and higher field regimes are fitted for the magnetic field below 1 T and above 6 T, respectively. The best fitting gives $\sqrt{a_1} \approx 6.89 \times 10^4$ cm$^2$V$^{-1}$s$^{-1}$ and $b_2 \approx 2270$ cm$^2$V$^{-1}$s$^{-1}$. Actually, there is still quadratic contribution left in the higher field regime as indicated in Figure S4. The fitting of the MR curves above 6 T with an additional quadratic term gives the temperature dependence of both components, as shown in Figure 4c. Obvious difference can be resolved between the linear and the quadratic components. The amplitude of the linear component is much larger than that of the quadratic component and shows a different temperature dependence. The linear component approaches a maximum around 200 K similar to that in Figure 4a. The obtained amplitudes are in the same order of magnitude as the fitting coefficients in eq 2. There are two distinct temperature regimes with a boundary around 200 K. Similar behavior can be observed from the temperature evolution of the critical field $B_c$, defined as the crossing point of the quadratic and linear fitting curves in Figure 4b. From Figure 4d, we can see that $B_c$ increases almost linearly with increasing temperature below 200 K. If we attribute this behavior to the competition between Zeeman energy $g^* \mu_B B$ and thermal energy $k_B T/2$,[2] where $g^*$ is an effective Lande factor, $\mu_B$ is Bohr magnetron and $k_B$ is

Boltzmann constant, we can obtain $g^*\approx 148$ from the linear fitting below 200 K, which is much higher than the literature reported values of α-Sn and InSb.[53] This may be related to the unusual MR behaviors and also indicate the MR properties should not be simply attributed to α-Sn or InSb. Another possibility is the Zeeman energy contributes only a small part to the linear component, and there may be other energies proportional to $B$ dominating the linear component. Above 200 K, $B_c$ starts to decrease with increasing temperature, further implying the change of the dominant mechanism at higher temperatures.

Zeeman splitting and orbital effect can induce peaks on the MR curve as suggested by Zheng et al.[41] Figure 5 shows the Hall resistance curves at different temperatures. The inset of Figure 5a shows the electron concentrations calculated by linear fitting of the Hall resistance below 1 T. The electron concentration stays nearly constant at low temperatures and increases obviously near room temperature. This supports the assumption of different conducting channels in two temperature regimes induced by thermally excited charge carriers. If we assume the distance between the two voltage contacts to be 2 mm, and the sample width to be 5 mm, we can estimate the mobility at 70 K to be about $6.74\times 10^4$ cm$^2$/Vs, close to $\sqrt{a_1}$ obtained previously. Usually $a_1 = \zeta\mu_H^2$, with the factor $\zeta \leq 1$ depending on the scattering mechanism, but there should not be difference in the magnitude compared to $\mu_H^2$. Considering the inhomogeneous electric field in the sample plane due to the point contacts, which leads to a larger measured resistivity and smaller $\mu_H$, the agreement between $\sqrt{a_1}$ and $\mu_H$ is reasonable. The Hall resistance shows an kinks around 2 T, indicating changes in either carrier concentration or type. The kinks are coincident with the MR peak position as shown in Figure 5b, revealing that the MR peak around 2.5 T may be related to the carrier modulation. This exactly reflects the modification effect of magnetic field on the Fermi surface. The orbital effect can shift LLs and introduce extra carriers,[41] and the competition between Zeeman energy and thermal energy results in the shift of this MR peak with increasing temperature as shown in Figure 4a. The Hall data comparison between the InSb substrate and the α-Sn/InSb heterostructure in Figure S5 also supports the modification effects of the α-Sn films.

In conclusion, we have grown high-quality epitaxial α-Sn films on InSb substrates and confirmed their thermal stability with coherent interfaces at room temperature. We take the α-Sn/InSb heterostructure as a whole and the effects of the α-Sn films on the transport properties are demonstrated by the comparison with the bare InSb substrate. A large MR over 450,000% at 14 T has been observed on the α-Sn/InSb heterostructures. From the thickness and angle dependent measurements, we believe that at least there is a quasi-2D Fermi surface for the large MR superimposed with a 3D Fermi surface for the oscillations. The MR magnitude shows a two-temperature-regime behavior. The drop of the linear MR near room temperature is possibly caused by the excitation of charge carriers in the InSb substrate. Extra carriers can also be introduce by the magnetic fields induced modifications on the Fermi surface. The exact reason why the α-Sn films has unusual effects on the MR properties need further investigations, however, our analysis may provide a reference for future researches on α-Sn films and studies on other topological material/semiconductor heterostructure system.

Specifically, the large MR persisting to near room temperature of α-Sn/InSb heterostructures is promising in applications of magnetic sensor or memory devices.

**Supporting information**

The Supporting Information is available free of charge on the ACS publication website.

Additional details on the surface morphology of α-Sn and β-Sn, aperiodicity of the oscillations, angle dependent MR of samples with different thicknesses, derivative of temperature dependent MR, and Hall resistance of the α-Sn sample and the InSb substrate at 2 K.

**Acknowledgements**

The authors acknowledge the support from the National Key R&D Program of China (2018YFA0306200, 2017YFA0303702), the National Natural Science Foundation of China (NSFC) (Grant No.51732006, No.11890702, No.51721001, No.91750101, No.21733001, No.51861145201).

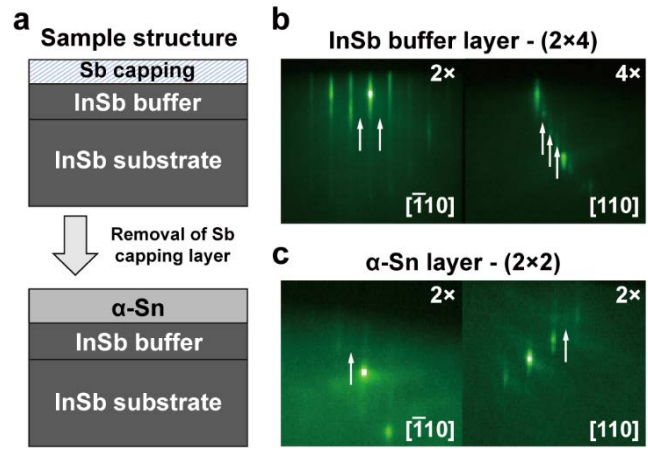

Figure 1. The two-step growth of the α-Sn samples. (a) The schematic layer structures of the α-Sn samples. The top structure shows the InSb template prepared in the III-V MBE system before the α-Sn growth. An amorphous Sb layer is used to protect the InSb surface from oxidation and contamination, and later removed by thermal evaporation before the α-Sn growth. The bottom structure shows the final α-Sn/InSb heterostructure. (b) The RHEED patterns of the InSb surface after oxide desorption showing a (2×4) surface reconstruction. (c) The RHEED patterns of the α-Sn films showing a (2×2) surface reconstruction.

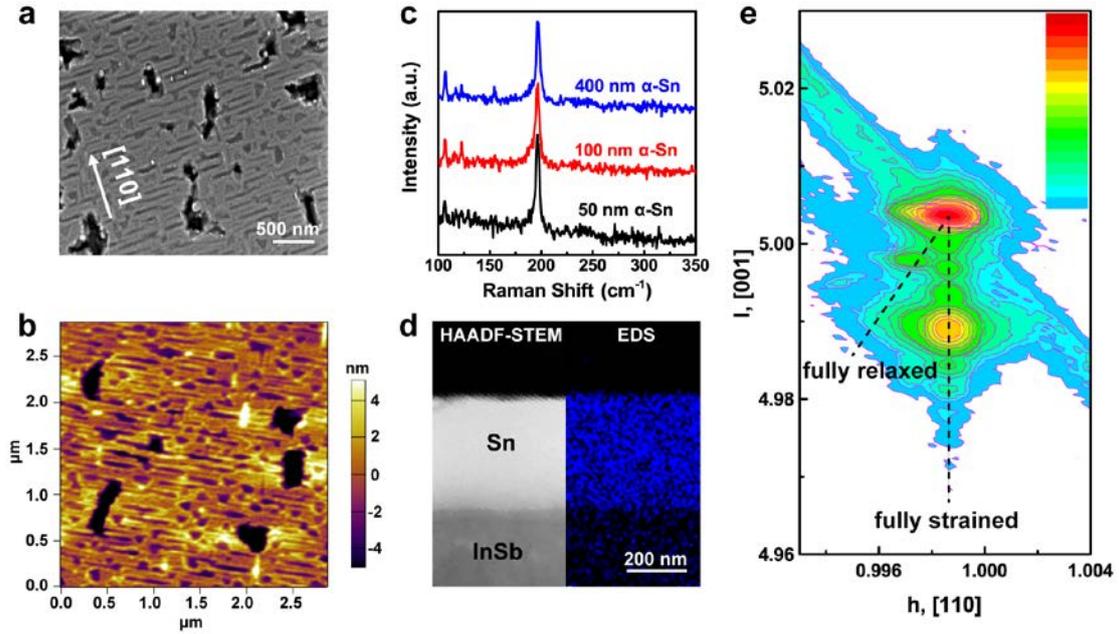

Figure 2. Structural characterizations of the α-Sn samples. (a) and (b) show the surface morphology of a 100 nm sample measured by SEM and AFM. (c) The Raman spectra of three α-Sn samples with difference thicknesses. The Raman shift of α-Sn around 196 cm$^{-1}$ can be clearly seen. (d) The cross-sectional HAADF-STEM (left) and corresponding EDS (right) images of a 400 nm α-Sn sample. The Sn signals are colored by blue. (e) The asymmetric RSM on the (115) plane of a 200 nm sample. The α-Sn film and the InSb substrate have the same in-plane reciprocal lattice constants, indicating the fully-strained state of the α-Sn film.

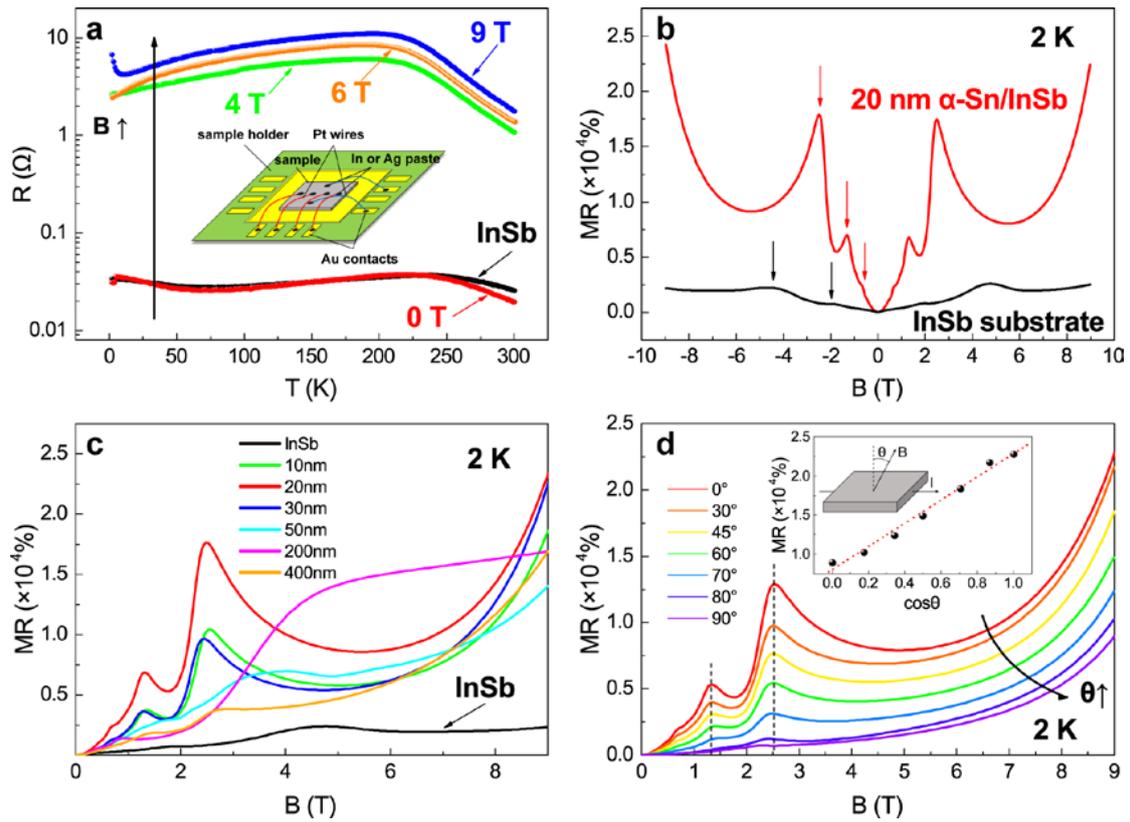

Figure 3. The MR properties of the α-Sn samples. (a) Temperature dependent resistance measured at 0, 4, 6 and 9 T on a 20 nm α-Sn/InSb heterostructure. The inset shows the four-probe configuration for the electrical measurements. (b) The MR of the α-Sn/InSb heterostructure compared to the InSb substrate at 2 K. The arrows denote the oscillation peaks on the MR curves. (c) The MR curves of a series of α-Sn/InSb heterostructures varying the α-Sn thickness. (d) The angle dependent MR curves of the 20 nm α-Sn/InSb heterostructure at 2 K. The dashed lines indicate the oscillation peak positions. The inset shows the MR as a function of cosθ at 9 T, and the red dashed line is the linear fitting of MR.

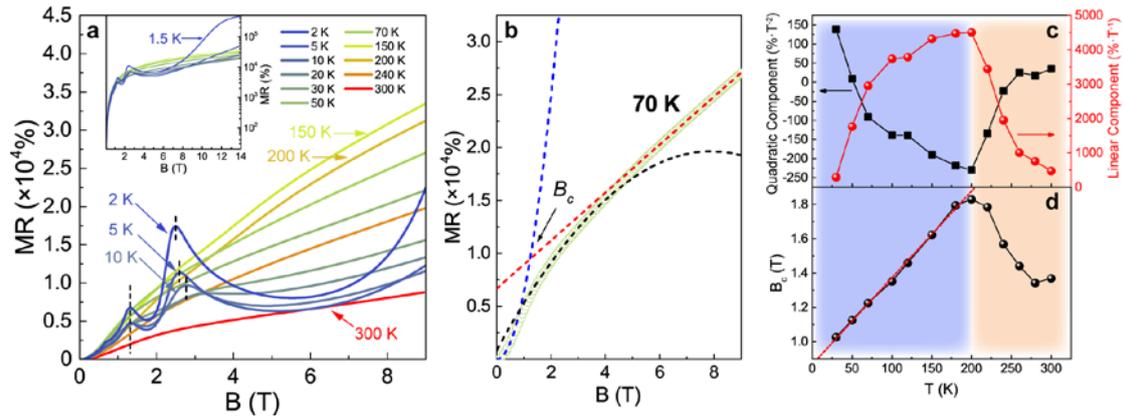

Figure 4. The temperature dependent MR properties. (a) The temperature dependent MR curves of the 20 nm α-Sn/InSb heterostructure. The inset shows the MR measured with 14 T at 1.5 K, yielding a 450,000% MR. (b) The fittings of the MR curve at 70 K. The $B_c$ is defined as the crossing point of the fitting curves of the higher and lower field regimes. (c) Temperature dependent quadratic and linear components obtained from fitting above 6 T with an extra quadratic term. (d) Temperature dependent critical magnetic field $B_c$ defined in (b).

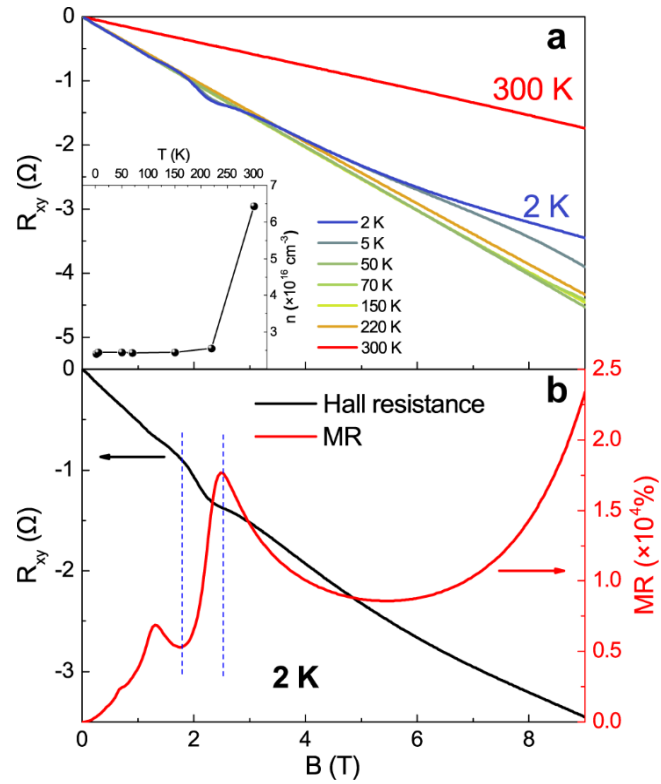

Figure 5. Hall effect measurements. The Hall resistance have been averaged using the data obtained in positive and negative magnetic fields. (a) The Hall resistance of the 20 nm α-Sn/InSb heterostructure at different temperatures. The inset shows the carrier concentrations at different temperatures calculated by fitting the Hall resistance curves below 1 T. (b) The Hall resistance and MR of the 20 nm α-Sn/InSb heterostructure at 2 K. Blue dashed lines indicate coincidence between the kinks on the Hall resistance curve and the extreme points on the MR curve.